\documentstyle[12pt,epsfig]{article}
\def\lsim{\raise0.3ex\hbox{$<$\kern-0.75em\raise-1.1ex\hbox{$\sim$}}}
\def\gsim{\raise0.3ex\hbox{$>$\kern-0.75em\raise-1.1ex\hbox{$\sim$}}}

\def\beq{\begin{equation}}
\def\eeq{\end{equation}}
\def\bea{\begin{eqnarray}}
\def\eea{\end{eqnarray}}
\def\bq{\begin{quote}}
\def\eq{\end{quote}}

\newcommand{\rr}{\mbox{\boldmath $r$}}

\newcommand{\rp}{\mbox{\boldmath $p$}}

\def\gappeq{\mathrel{\rlap {\raise.5ex\hbox{$>$}}
{\lower.5ex\hbox{$\sim$}}}}

\def\lappeq{\mathrel{\rlap{\raise.5ex\hbox{$<$}}
{\lower.5ex\hbox{$\sim$}}}}

\def\Toprel#1\over#2{\mathrel{\mathop{#2}\limits^{#1}}}

\newcommand{\rb}{\mbox{\boldmath $b$}}
\newcommand{\rk}{\mbox{\boldmath $k$}}

\newcommand{\rkn}{\mbox{$k$}}

\begin{document}
\pagestyle{empty}
%\begin{flushright}
%{CERN-TH/2001-265}\\
%hep-ph/th number??\\
%\end{flushright}
%\vspace*{5mm}
\begin{center}
{\bf THE QCD POMERON IN ULTRAPERIPHERAL HEAVY ION COLLISIONS: \\ III. PHOTONUCLEAR PRODUCTION OF HEAVY-QUARKS}
\\
\vspace*{1cm}
 V.P. Gon\c{c}alves $^{1}$, M.V.T. Machado  $^{1,\,2}$\\
\vspace{0.3cm}
{$^{1}$ Instituto de F\'{\i}sica e Matem\'atica,  Universidade
Federal de Pelotas\\
Caixa Postal 354, CEP 96010-090, Pelotas, RS, Brazil\\
$^{2}$ \rm High Energy Physics Phenomenology Group, GFPAE,  IF-UFRGS \\
Caixa Postal 15051, CEP 91501-970, Porto Alegre, RS, Brazil}\\
\vspace*{1cm}
{\bf ABSTRACT}
\end{center}

\vspace*{1cm} \noindent

\vspace*{1.5cm} \noindent \rule[.1in]{17cm}{.002in}

\vspace{-4.2cm} \setcounter{page}{1} \pagestyle{plain} We
calculate the photonuclear production of heavy quarks in
ultraperipheral heavy ion collisions. The integrated cross section
and the rapidity distribution are computed employing sound high
energy QCD formalisms as the collinear and semihard approaches as
well as the saturation model. In particular, the color glass
condensate (CGC)  formalism is also considered  using a  simple
phenomenological parameterization for the color field correlator
in the medium, which allow us to obtain more  reliable  estimates
for charm and bottom production at
 LHC energies.

\vspace{0.3cm}

\section{Introduction}
The recent results from RHIC suggest that relativistic heavy ion
collisions  at high energies probe QCD in the non-linear regime of
high parton density \cite{kharzlevin}. In such a regime  the
growth of parton distributions should saturate, possibly forming a
color glass condensate \cite{iancu} (for a pedagogical
presentation  see Refs. \cite{mcllec,jamaliancu}), which is
characterized by a bulk momentum scale $Q_s$. In particular, the
RHIC data on the multiplicity distribution of the produced hadrons
as a function of centrality, rapidity, and collision energy
appears to be consistent with the CGC predictions.  However, there
are still a number of open questions, mainly associated to the
fact that other models based on different assumptions reasonably
describe the same set of experimental data
\cite{eskqm,Armestopaj}. For instance, the recent RHIC data
indicate  the lack of suppression in  the high-$p_T$ spectra of
charged hadrons produced in $d+Au$ collisions \cite{rhicda}, in
contrast with the initial expectation of the CGC formalism
\cite{kharzlevin}. The main uncertainty present in those analysis
is directly connected with the poor knowledge of the initial
conditions of the heavy ion collisions. Theoretically, the early
evolution of these nuclear collisions is governed by the dominant
role of gluons \cite{vni}, due to their large interaction
probability and the large gluonic component in the initial nuclear
wave functions. Consequently, a systematic measurement of the
nuclear gluon distribution is of fundamental interest to
understand the parton structure of nuclei, determine the initial
conditions of the QGP and constraint the QCD dynamics at high
energies.

In the last years, there has been a lot of interest in the
description of electron-nucleus collisions at high energies, with
particular emphasis in the behavior of the nuclear structure
functions \cite{nucsfs} and their logarithmic slope
\cite{vicslope} at small value of the Bjorken $x$ variable,
obtaining predictions which agree with the scarce experimental
data. Moreover, the high energy heavy quark photoproduction on
nuclei targets has been studied in detail in Refs.
\cite{Goncalves:2003kp,Goncalves:2003zy}, considering the
available several scenarios for the QCD dynamics at high energies.
The results of those analysis show that future  electron-nucleus
colliders at HERA and RHIC \cite{HERAeA,raju}, probably could
determine whether parton distributions saturate and  constraint
the behavior of the nuclear gluon distribution in the full
kinematical range. However, until these colliders become reality
we need to consider alternative searches  in the current and/or
scheduled  accelerators which allow us to constraint the QCD
dynamics. Here, we analyze the possibility of using
ultraperipheral heavy ion collisions as a photonuclear collider
and study the heavy quark production  assuming distinct approaches
for the QCD evolution.

 The analysis of heavy quark production  in ultraperipheral heavy ion
collisions has been proposed many years ago \cite{perihe1},
improved in  Refs. \cite{perihe2,perihe3,perihe4} and  recently
revisited in Refs. \cite{gelis,vicber,ramonaklein}. The most of
these approaches calculates the cross section assuming the
validity of the collinear factorization, where the cross sections
involving incoming hadrons are given,  at all orders, by the
convolution of intrinsically non-perturbative, but universal,
quantities - the parton densities, with perturbatively calculable
hard matrix elements, which are process dependent. In this
approach  all partons involved are assumed to be on mass shell,
carrying only longitudinal momenta, and their transverse momenta
are neglected in the QCD matrix elements. However,  in the large
energy (small-$x$) limit,  the effects of the finite transverse
momenta of the incoming partons become important, and the
factorization must be generalized, implying that the cross
sections are now $k_{\perp}$-factorized into an off-shell partonic
cross section and a  $k_{\perp}$-unintegrated parton density
function ${\cal{F}}(x,k_{\perp})$, characterizing the
$k_{\perp}$-factorization  approach \cite{CCH,CE,GLRSS}.  The
function $\cal{F}$ is obtained as a solution  of the   evolution
equations associated to the dynamics that governs the QCD at high
energies.   Here, we estimate, for the first time, the total cross
section and the rapidity dependence of the nuclear photoproduction
of heavy quarks in ultraperipheral heavy ion collisions,
considering the $k_{\perp}$-factorization approach and distinct
nuclear unintegrated gluon distributions. Moreover, we extend the
previous study for heavy quark production in the color glass
condensate formalism considering a realistic photon flux and a
phenomenological dipole cross section which is energy dependent,
allowing to obtain reliable estimates  for the rapidity
distribution within this  formalism.  For comparison, we also
present the predictions for the cross section from the collinear
factorization approach.

This paper is organized as follows. In  next section we present a
brief review of ultraperipheral heavy ion collisions and write
down the main formulas describing the photonuclear process in
these collisions. In Section \ref{models} we discuss some sound
models for the heavy quark photoproduction in the collinear and
$k_{\perp}$-factorization approaches. Moreover, we discuss our
main assumptions  in order to extend the previous results  on
color glass condensate formalism. Finally, in Section
\ref{discussion} we present our results for the total cross
section and rapididity distribution of charm and bottom production
for the LHC energies.

\section{Ultraperipheral relativistic heavy ion collisions}

In heavy ion  collisions the large number of photons coming from
one of the colliding nuclei  will  allow to study photoproduction,
with energies $W_{\gamma N}$ reaching to  950 GeV for the LHC. The
photonuclear cross sections are given by the convolution between
the photon flux from one of the nuclei and the cross section for
the scattering photon-nuclei. The photon flux is given by the
Weizsacker-Williams method \cite{BaurPR}. The flux from a charge
$Z$ nucleus a distance $b$ away is
\begin{eqnarray}
\frac{d^3N\,(\omega,\,b^2)}{d\omega\,d^2b}= \frac{Z^2\alpha_{em}\eta^2}{\pi^2 \,\omega\, b^2}\, \left[K_1^2\,(\eta) + \frac{1}{\gamma_L^2}\,K_0^2\,(\eta) \right] \,
\label{fluxunint}
\end{eqnarray}
where $\gamma_L$ is the Lorentz boost  of a single beam and $\eta
= \omega b/\gamma_L$; $K_0(\eta)$ and  $K_1(\eta)$ are the
modified Bessel functions. The requirement that  photoproduction
is not accompanied by hadronic interaction (ultraperipheral
collision) can be done by restricting the impact parameter $b$  to
be larger than twice the nuclear radius, $R_A=1.2 \,A^{1/3}$ fm.
Therefore, the total photon flux interacting with the target
nucleus is given by Eq. (\ref{fluxunint}) integrated over the
transverse area of the target for all impact parameters subject to
the constraint that the two nuclei do not interact hadronically.
An analytic approximation for $AA$ collisions can be obtained
using as integration limit $b>2\,R_A$, producing
\begin{eqnarray}
\frac{dN\,(\omega)}{d\omega}= \frac{2\,Z^2\alpha_{em}}{\pi\,\omega}\, \left[\bar{\eta}\,K_0\,(\bar{\eta})\, K_1\,(\bar{\eta})+ \frac{\bar{\eta}^2}{2}\,\left(K_1^2\,(\bar{\eta})-  K_0^2\,(\bar{\eta}) \right) \right] \,
\label{fluxint}
\end{eqnarray}
where $\bar{\eta}=2\omega\,R_A/\gamma_L$. The final  expression
for the production of heavy quarks in ultraperipheral heavy ion
collisions is then given by,
\begin{eqnarray}
\sigma_{AA \rightarrow Q\overline{Q}X}\,\left(\sqrt{S_{\mathrm{NN}}}\right) = \int \limits_{\omega_{min}}^{\infty} d\omega \, \frac{dN\,(\omega)}{d\omega}\,\, \sigma_{\gamma A \rightarrow Q\overline{Q}X} \left(W_{\gamma A}^2=2\,\omega\sqrt{S_{\mathrm{NN}}}\right)\,
\label{sigAA}
\end{eqnarray}
where $\omega_{min}=M_{Q\overline{Q}}^2/4\gamma_L m_p$ and
$\sqrt{S_{\mathrm{NN}}}$ is  the c.m.s energy of the
nucleus-nucleus system. The Lorentz factor for LHC is
$\gamma_L=2930$, giving the maximum c.m.s. $\gamma N$ energy
$W_{\gamma A} \lappeq 950$ GeV.  It is worth mentioning that the
difference between the complete numeric and the analytical
calculation presented above  for the photon flux is less than 15
\% for the most of the purposes \cite{BaurPR}.

Before considering  the distinct  models for the photon-nucleus
cross section, it is interesting to determine the values of $x$
which will be probed in ultraperipheral heavy ion collisions. The
Bjorken $x$ variable is given by $x = (M_{Q\overline{Q}}/2p)
e^{-y}$, where $M_{Q\overline{Q}}$ is the invariant mass of the photon-gluon
system and $y$ the center of momentum rapidity. For Pb + Pb
collisions at LHC energies the nucleon momentum is equal to
$p=2750$ GeV; hence $x = (M_{Q\overline{Q}}/5500 \, {\rm GeV})
e^{-y}$. Therefore, the region of small mass and large rapidities
probes directly the high energy (small $x$) behavior of the QCD
dynamics present in the $\gamma \, A$ cross section. This
demonstrates that ultraperipheral heavy ion collisions at LHC
represents a very good tool to constraint the high energy regime
of the QCD dynamics, as already verified for two-photon processes
\cite{vicmag1,vicmag2,vicmag3}.

\section{Models for nuclear heavy quark production}
\label{models}

For our further analysis on photonuclear  production of heavy
quarks we will consider distinct available high energy approaches.
First, we take into account the usual collinear approach, where
the production cross section is driven by the collinear gluon
distribution on the nuclei. This one contains a lot of information
about nuclear shadowing, EMC, and anti-shadowing effects. Second,
the $\rk_{\perp}$-factorization formalism is introduced, where the
relevant quantity is now the nuclear unintegrated gluon
distribution. For this purpose, we analyze two simple
parameterization for it which are consistent with data description
on inclusive and diffractive DIS. Finally, we take into account
the color glass condensate formalism, where the scattering process
is viewed as the interaction of the probe particles with the
strong nuclear color field treated in a classical approximation.
The main quantity is the correlator of two Wilson lines, which is
related to the dipole cross section and at  lowest order has no
energy dependence. We  present a simple phenomenological
parameterization which introduces higher orders corrections to the
classical approximation and allow us produce more realistic
estimates for the cross section.

\subsection{The collinear approach}

 In hard photon-hadron interactions the
photon can behave as a pointlike particle in the so-called direct
photon processes or it can act as a source of partons, which then
scatter against partons in the hadron, in the resolved photon
processes (For a recent review see Ref. \cite{nisius}). Resolved
interactions stem from the photon fluctuation into a quark-antiquark
state or a more complex partonic state, which  are embedded in the
definition of the photon structure functions. Recently, the
contribution of resolved photon processes in ultraperipheral heavy
ion collisions was discussed in Ref. \cite{vicber} and studied in
detail in Ref. \cite{ramonaklein}. One of the main results is that
at LHC, these contributions are $\approx$ 15 and 20 $\%$ of the
total charm and bottom photoproduction cross sections,
respectively, comparable to the shadowing effect. Here, we will
consider only the direct photon contribution.

At high energies the main subprocess occurring when the photon
probes the structure of the nucleus is the photon-gluon fusion
producing the heavy quark pair. It can be  described  through
perturbative QCD, with the cross section given in terms of the
convolution between the elementary cross section for the
subprocess $\gamma g \rightarrow Q \overline{Q}$ and the
probability of finding a gluon inside the nucleus, namely the
nuclear gluon distribution. In this collinear approach the  heavy
quark  photoproduction cross section is given by
\begin{eqnarray}
\sigma_{\gamma A\rightarrow Q\overline{Q}}\, (W_{\gamma A}) &=&
\int_{4m_Q^2}^{W_{\gamma A}^2}\,dM_{Q\overline{Q}}^2
 \,  \frac{d\sigma_{\gamma g \rightarrow Q\overline{Q}}}{dM_{Q\overline{Q}}^2}\, g_A(x,\mu^2)\,\,,
    \label{sigpho}
\end{eqnarray}
where the quantity $d\sigma_{Q\overline{Q}}/dM^2_{Q\overline{Q}}$
is calculable perturbatively, $M_{Q\overline{Q}}$ is the invariant
mass of the heavy quark  pair with
$x=M^2_{Q\overline{Q}}/W_{\gamma A}^2$.  The c.m.s  energy of the
$\gamma A$ system is labelled $W_{\gamma A}$ and  $g_A(x, \mu^2)$
is the gluon density inside the nuclear medium, with $\mu$ being
the factorization scale   and $m_Q$  the heavy quark mass. For our
purpose here we will use $\mu^2=4m_Q^2$, with $m_c = 1.5$ GeV and
$m_b = 4.5$ GeV.  The differential cross section in leading order
is given by \cite{Gluck78}
\begin{eqnarray}
\frac{d\sigma_{\gamma g\rightarrow Q\overline{Q}}}
{dM_{Q\overline{Q}}} = \frac {4\pi\,\alpha_{em}\,\alpha_s(\mu^2)\,e_Q^2}
{M^2_{Q\overline{Q}}} \left[
(1+\beta +\frac{1}{2}\beta^2)\ln(\frac{1+\sqrt{1-\beta}}
{1-\sqrt{1-\beta}})  -(1+\beta) \sqrt{1-\beta}\right] \,\,,
\label{integ}
\end{eqnarray}
where $e_Q$ is the heavy quark charge and
$\beta=4\,m_Q^2/M_{Q\overline{Q}}^2$. In our further  calculation
on the collinear approach one considers that $xg_A(x, Q^2) =
R_g(x,Q^2) \times xg_N(x,Q^2)$, where $R_g$ parameterize the
medium effects as proposed in Ref. \cite{EKS} and $xg_N$  is the
nucleon gluon distribution  given by the GRV98(LO)
parameterization \cite{grv98}. It is worth mentioning that
different choices for the factorization scale and quark mass
produce distinct overall normalization to the total cross section
at photon-nucleus  and ultraperipheral nucleus-nucleus
interactions.  For details see Ref. \cite{Goncalves:2003zy}, where
the heavy quark photoproduction at eRHIC and THERA energies has
been discussed. In the Section \ref{discussion} we discuss the
dependence of our results on the choice of the quark mass and
parton distribution parameterization.

\subsection{The $\rk_{\perp}$-factorization formalism}

In the $k_{\perp}$-factorization (or semihard) approach, the
relevant QCD diagrams are considered with the virtualities and
polarizations of the initial partons, carrying information on
their transverse momenta. The scattering processes are described
through the convolution of off-shell matrix elements with the
unintegrated parton distribution, ${\cal F}(x,\rk_{\perp})$ (see
\cite{smallx} for a review).  The latter can recover the usual
parton distributions in the double logarithmic limit  by its
integration over the transverse momentum of the $\rk_{\perp}$
exchanged gluon. The gluon longitudinal momentum fraction is
related to the c.m.s. energy, $W_{\gamma \,A}$,  in the heavy
quark photoproduction case as $x=4m_{Q}^2/W_{\gamma \,A}^2$, as in
the collinear case. The cross section for the heavy-quark
photoproduction process is given by the convolution of the
unintegrated gluon function with the off-shell matrix elements
\cite{smallx,semih,Mariotto_Machado,Timneanu_Motyka}.  Considering
only the direct component of the photon we have that
$\sigma_{tot}^{phot}$ reads  as \cite{Mariotto_Machado},
\begin{eqnarray}
&\sigma_{tot}^{phot} (W_{\gamma A})& =
  \frac{\alpha_{em}\,e_Q^2}{\pi}\, \int\, dz\,\,d^2 \rp_{1\perp} \, d^2\rk_{\perp} \, \frac{\alpha_s(\mu^2)\,{\cal F}(x,\rk_{\perp}^2; \,\mu^2)}{\rk_{\perp}^2}\nonumber \\
&&\!\!\!\!\!\!\times
 \left\{ [z^2+ (1-z)^2]\,\left( \frac{\rp_{1\perp}}{D_1} + \frac{(\rk_{\perp}-\rp_{1\perp})}{D_2} \right)^2 +   m_Q^2 \,\left(\frac{1}{D_1} + \frac{1}{D_2}  \right)^2  \right\}\,, \label{sigmakt}
\end{eqnarray}
where $D_1 \equiv \rp_{1\perp}^2 + m_Q^2$ and $D_2 \equiv (\rk_{\perp}-\rp _{1\perp})^2 + m_Q^2$. The transverse momenta of the heavy quark (antiquark) are denoted by $\rp_{1\perp}$ and $\rp_{2\perp}= (\rk_{\perp}-\rp _{1\perp})$, respectively. The heavy quark longitudinal momentum fraction is labeled by $z$. The scale $\mu$ in the strong coupling constant in general is taken to be equal to the gluon virtuality,  in close connection with the BLM scheme \cite{BLM}. Here, we will use the prescription  $\mu^2=\rk_{\perp}^2 + m_{Q}^2$.

In order to perform a phenomenological analysis within the
$k_{\perp}$-factorization approach, in the following we use two
distinct  parameterizations for the unintegrated gluon
distribution (For details see Ref. \cite{Goncalves:2003zy}).
First, one considers the derivative of the collinear nuclear gluon
parton distribution function, quite successful in the proton case
and tested in the nuclear case in Ref. \cite{Goncalves:2003zy}. It
simply reads  as,
\begin{eqnarray}
 {\cal F}_{\mathrm{nuc}}\,(x,\,\rk_{\perp}^2;\,A) =  \frac{\partial\, xG_{A}(x,\,\rk_{\perp}^2)}{\partial \ln \rk_{\perp}^2}\,,
\label{nucugf}
\end{eqnarray}
where $xG_{A}(x,Q^2)$ is the nuclear gluon distribution, which was
taken from the  EKS parameterization \cite{EKS} for the medium
effects and the GRV94(LO) for the nucleon parton distribution
\cite{grv94}. The latter choice is supported  by the good
description of heavy quark photoproduction in the full kinematical
region \cite{Goncalves:2003zy}. As a consequence, with this
procedure we include in our calculations the medium effects
(shadowing, antishadowing, EMC and Fermi motion effects) estimated
by that parameterization. Moreover, we emphasize that this nuclear
gluon distribution is solution of the DGLAP evolution equations,
which is associated to a linear dynamics that  does not consider
dynamical saturation effects.

The second parameterization is given by the model introduced  in
Ref. \cite{armesto}, which provides  an extension of the $ep$
saturation model through  Glauber-Gribov formalism. In this model
the cross section for the  heavy quark photoproduction on nuclei
targets is given by  \cite{armesto,Goncalves:2003kp}
\begin{eqnarray}
\sigma_{tot}^{\gamma\,A} (W, A)  = \int_0^1
dz\, \int d^2\rr \, |\Psi_{T} (z,\,\rr,\,Q^2=0)|^2 \, \sigma_{dip}^{\mathrm{A}}
(\tilde{x},\,\rr^2,A)\,,
\label{sigmaphot}
\end{eqnarray}
where the transverse wave function is known  (See e.g. Ref.
\cite{predazzi}). As $|\Psi_L|^2 \propto Q^2$, the longitudinal
piece does not contribute for $Q^2 = 0$. The nuclear dipole cross
section is given by,
\begin{eqnarray}
\sigma_{dip}^{\mathrm{A}} (\tilde{x}, \,\rr^2, A)  = \int d^2b \,\, 2
\left\{\, 1- \exp \left[-\frac{1}{2}\,A\,T_A(b)\,\sigma_{dip}^{\mathrm{p}} (\tilde{x}, \,\rr^2)  \right] \, \right\}\,,
\label{sigmanuc}
\end{eqnarray}
where $b$ is the impact parameter of the center  of the dipole
relative to the center of the nucleus and the integrand gives the
total dipole-nucleus cross section for fixed impact parameter. The
nuclear profile function is labelled by $T_A(b)$, which will be
obtained from the 3-parameter Fermi distribution for the nuclear
density \cite{Devries}.   The parameterization for the dipole
cross section takes the eikonal-like form,
$\sigma_{dip}^{\mathrm{p}} (\tilde{x},\,\rr^2)  =  \sigma_0 \,[\,
1- \exp \left(-Q_s^2(\tilde{x})\,\rr^2/4 \right) \,]$, where one
has  used the parameters from \cite{GBW}, which include the charm
quark with mass $m_c=1.5$ GeV and the definition $ \tilde{x}=(Q^2
+ 4\,m_Q^2)/W_{\gamma A}^2$. The saturation scale $Q_s^2(x)=
\left(x_0/x \right)^{\lambda}$ GeV$^2$, gives the onset of the
saturation phenomenon to  the process.

The  equation above sums up all the multiple  elastic rescattering
diagrams of the $q \overline{q}$ pair and is justified for large
coherence length, where the transverse separation $r$ of partons
in the multiparton Fock state of the photon becomes as good a
conserved quantity as the angular momentum, namely the size of the
pair $r$ becomes eigenvalue of the scattering matrix.  The
corresponding unintegrated gluon distribution can be recovered
from a Bessel-Fourier transform to the momentum representation
\cite{armesto},
\begin{eqnarray}
 {\cal F}_{\mathrm{nuc}}\,(x,\,\rk_{\perp}^2, b)  =   \frac{N_c}{\pi^2 \alpha_s} \,\left( \frac{\rk_{\perp}^2}{Q_s^2}\right)\, \sum_{m=1}^{\infty} \,\sum_{n=0}^{m}\,\frac{\left(-\frac{1}{2}\,A\,T_A(b)\,\sigma_0\right)^m}{m\, !}\,C^{n}_{m}\,\frac{(-1)^n}{n} \exp\,\left( -\frac{\rk_{\perp}^2}{n\,Q_s^2}\right)\,
\end{eqnarray}
which depends on the transverse momentum $\rk_{\perp}$  through
the scaling variable $\tau \equiv  \rk_{\perp}^2/Q_s^2$. The
unintegrated gluon vanishes asymptotically at $ \rk_{\perp}^2
\rightarrow 0,\,\infty$ and its maximum can be identified with the
saturation scale $Q_{s\,A}(x)$ \cite{armesto,Armesto_Braun}. The
model recovers the original one for the proton case, taking $A=1$
and the normalization condition $\int d^2b\,T_A(b)=1$.

In Ref. \cite{Goncalves:2003kp}, it has verified that  the
resummation at  the proton level is less sizeable in the final
results at nuclear level. Therefore, this fact allow us take just
the color transparency behavior on the dipole nucleon cross
section. Hence, in such a particular case one can  compute
analytically the unintegrated gluon distribution, which  is
expressed as
\begin{eqnarray}
 {\cal F}_{\mathrm{nuc}}\,(x,\,\rk_{\perp}^2, b)= \frac{N_c}{2\,\alpha_s\,\pi^2}\,\left(\frac{\rk_{\perp}^2}{Q_{s\,A}^2(x)} \right)\, \exp \left(-\frac{\rk_{\perp}^2}{Q_{s\,A}^2(x)}  \right)\,,
\label{gluonnuc}
\end{eqnarray}
where $Q_{s\,A}^2(x)=\frac{1}{2}A\,T_A(b)\,\sigma_0\,Q_s^2(x)$  define the nuclear saturation scale. Such an approximation is justified in the heavy quark case, which is dominated by small dipole configurations  (large transverse momentum $\rk_{\perp}^2 \simeq m_Q^2$).  It is clear that Eq.  (\ref{gluonnuc}) presents a  scaling pattern  on the variable $\tau=\rk_{\perp}^2/Q_{s A}^2$, which implies scaling on $\tau$ in the nuclear heavy quark production. Recently, this feature has been  shown also in the nucleon case \cite{charmletter}.

Some comments are in order here. In our analysis we have
disregarded higher-order Fock states in the photon wave function
and considered only the evolution in the dipole cross section. In
principle, at leading order and for inclusive process this is a
reasonable approximation. However, for diffractive processes, for
example,  the $q\overline{q} g$ component cannot be disregarded.
The recent results for the NLO corrections for the impact factor
\cite{bartelsnlo} will allow to verify the validity of the color
dipole approach at higher orders. Other point which deserves
discussion is that we have assumed the validity of the
$k_{\perp}$-factorization in photon-nucleus interactions. At high
energies and intermediate densities, it is a reasonable
assumption, since the derivation of the $k_{\perp}$-factorization
presented in Ref. \cite{catani} can be directly extended for the
nuclear heavy quark photoproduction. However, for very  large
parton densities a breakdown of the $k_{\perp}$ (and collinear)
factorization is expected, mainly associated to the effects of the
non-linearity of the non-Abelian gluon field
\cite{iancu,gelisjam1,rajukran}. Finally, we also have disregarded
the resolved photon contribution in the semihard approach. For
completeness, lets quote the estimates for the resolved component
on the nucleon level. In Ref. \cite{Timneanu_Motyka} it gives a
contribution of order 20-30 \%, rising faster on energy (hence
important mostly at large rapidities)  than the direct photon component  and
being stable under different choices for  the unintegrated gluon
distributions and quark mass. Moreover, the resolved contribution
is somewhat higher than in the collinear approach due to the
non-zero transverse momentum transfer effect. We expect a similar
trend in photon-nucleus interactions.

\subsection{The Color Glass Condensate formalism}

At small $x$ and/or large $A$ one expects the transition of the
regime described by the linear dynamics (DGLAP, BFKL) (For a
review, see e.g. Ref. \cite{caldwell}), where only the parton
emissions are considered, to a new regime where the physical
process of recombination of partons becomes important in the
parton cascade and the evolution is given by a nonlinear evolution
equation. In this  regime    a Color Glass Condensate (CGC) is
expected to be formed \cite{iancu}, being characterized by the
limitation on the maximum phase-space parton density that can be
reached in the hadron/nuclear wavefunction (parton saturation) and
very high values of the QCD field strength $F_{\mu \nu} \approx
1/\sqrt{\alpha_s}$ \cite{mue}. The large values  of the gluon
distribution at saturation (large occupation number of the soft
gluon modes)  suggests the use of semi-classical methods, which
allow to describe the small-$x$ gluons inside a fast moving
nucleus by a classical color field. This color field is driven by
a classical Yang-Mills equation whose source term is provided by
faster partons.  When the energy further increases, the structure
of the classical field equations does not change, but only the
correlations of the color source. This change can be computed in
perturbation theory, and expressed as a functional renormalization
group equation for the weight function, in which the 'fast'
partons are integrated out in steps of rapidity and in the
background of the classical field generated at the previous step.
This approach  enables one to calculate cross section cross
sections in a high gluon density environment. Recently, this
approach has been applied for $eA$ \cite{rajumcl,gelisjam1}, $pA$
\cite{gelisjam1,dumitru,dumjam} and $AA$ processes
\cite{rajukran}.

In Refs. \cite{gelis} the heavy quark production  in
ultraperipheral heavy ion collisions has been analyzed in the
color glass condensate formalism. In particular, those authors
have considered the photon-nuclei interaction, taking into account
the electromagnetic interaction to lowest order in the coupling
and the interactions with the strong color background field to all
orders. The quantum evolution is not included in the calculations.
Their prediction for the rapidity ($y$) distribution of the heavy
quark (or antiquark) is given by
\begin{eqnarray}
\frac{d\sigma_{AA\rightarrow Q\overline{Q}X}}{dy}=\pi R_A^2\,\frac{N_c (Z\alpha_{em})^2\,e_q^2}{6\pi^4} \!\int\limits_{2R_A}^{\frac{\gamma_{L}}{m_Q}} \frac{d^2\rb}{\rb^2} \,\int\limits_{0}^{+\infty} d\rk_{\perp}^2\, C(\rk_{\perp})
\left\{ 1+ \frac{4(\rk_{\perp}^2-m_Q^2)}{\rkn_{\perp}\sqrt{\rk_{\perp}^2+4m_Q^2}}\,{\rm arcth}\,\frac{\rkn_{\perp}}{\sqrt{\rk_{\perp}^2+4m_Q^2}} \right\}\!,
\label{dsdy_cgc}
\end{eqnarray}
with $N_c$ being the color number and $e_q$ the  quark charge. The
color field correlator $C(\rk_{\perp})$ in the medium is given by
\begin{eqnarray}
C(\rk_{\perp})\equiv \int d^2\rr\, e^{\,i\,{\rk}_{\perp}\cdot{\rr}}\, e^{-B_2(\rr)} = \int d^2\rr\, e^{\,i\,{\rk}_{\perp}\cdot{\rr}}
\,\left<U(0)U^\dagger(\rr) \right>_{\rho},
\label{correlator}
\end{eqnarray}
with $\left< .. \right>_{\rho}$ representing the  average over all
the configurations of the color fields in the nucleus. The
unitarity matrix $U(\rr)$ contains the information related to the
interactions between the quark and the colored glass condensate
(classical color field of the nucleus) and is expressed in terms
of the color sources in the nucleus. Therefore, $C(\rk_{\perp})$
depends on the structure of the color sources describing the
target nucleus and describes the interactions of a high energy
probe with the target.

In Ref. \cite{gelis} the McLerran-Venugopalan model  for the
correlator  $C(\rk_{\perp})$ was considered. In this model the
function $B_2 (\rr)$ in  Eq. (\ref{correlator}) is  approximated
as follows
\begin{eqnarray}
B_2(\rr)\approx {{Q_{sA}^2\,\rr^2}\over{4\pi}}\ln\left({1\over{\rr\, \Lambda}}\right)\; ,
\label{B2-approx}
\end{eqnarray}
where the saturation scale $Q_{sA}$ at this classical  level does
not depend on the rapidity (energy) and $\Lambda$ is an infrared
cutoff related to the scale at which color neutrality occurs. In
principle, $\Lambda$ is at least as large as $\Lambda_{QCD}$. For
a saturated target, it can be probed that  color neutrality occurs
over transverse spatial scales as small as $1/Q_s$
\cite{Ferreiro:2002kv}. Lets summarize the results coming from
applying the approximation above and $\Lambda = \Lambda_{QCD}$ in
Eq. (\ref{dsdy_cgc}). Supposing $Q_{sA}/\Lambda_{QCD}=10$
($y=2.3$) one obtains $d\sigma/dy=355$ mb for charm and
$d\sigma/dy=11$ mb for bottom with the rapidity distribution being
flat on $y$.

In principle, extending the  previous calculation to include
quantum evolution is just a matter of changing the
$C(\rk_{\perp})$ present in the previous calculations by one that
has been calculated considering the evolution of the sources.
However, the general solution of the functional renormalization
group equation for the weight function is not known, but only
approximate solutions  in some limiting kinematical regimes have
been obtained \cite{Iancu:2002aq}.  In general, the quantum
corrections lead to a modification of the distribution function of
the hard sources when the energy increases, which implies a
rapidity dependence for the correlator  $C(\rk_{\perp})$.

In order  to go further and introduce an energy dependence on the
calculations one makes use of the fact that the function
$C(\rk_{\perp})$ is directly related to the Fourier transform of
the  dipole-nucleus total cross section, as follows
 \begin{eqnarray}
C\, (x,\rk_{\perp}) \equiv \frac{1}{2\pi R_A^2} \int d^2 \rr \, e^{\,i\,\rk_{\perp} \cdot \rr} \left[\sigma_{dip} (x,\rr \rightarrow \infty) - \sigma_{dip} (x,\rr)\right]\,\,.
\label{corel}
\end{eqnarray}
For this definition of the correlator, we have a direct relation
between this quantity and the unintegrated gluon density, given by
${\cal{F}}(x,\rk_{\perp}) = (3 R_A^2 / 8 \pi^2 \alpha_s)\,
\rk_{\perp}^2 \,C\, (x,\rk_{\perp})$. In our analysis we will
employ an educated guess for the dipole-nucleus cross section
\begin{eqnarray}
\sigma_{dip}^A (x,\rr) = 2\pi R_A^2\,\left[1 - \exp \left( - \frac{Q_{sA}^2 (x)\,\rr^2}{4} \right)\right]\,\,,
\label{sdip}
\end{eqnarray}
where  we assume $Q_{s\,A}^2 (x) = A^{1/3} \times Q_{s}^2 (x)$.
This model is inspired in the saturation model for the
dipole-proton  scattering proposed in Ref. \cite{GBW}, and
encodes the main properties of the high density  approaches,
namely  color transparency for small pair separations ($r
\rightarrow 0$) and saturation for large pair separations
($r\rightarrow \infty$). Moreover, the nuclear $A$-dependence
assumed for the saturation scale agrees with the recent analysis
for the scattering of a small dipole in a nuclear target in the
McLerran-Venugopalan model and fixed coupling BFKL dynamics
\cite{muea}. However, the energy dependence of the saturation
scale still is an open question. Recently,  several groups  have
studied the solutions of the Balitski-Kovchegov equation
\cite{qsx} (See also Ref. \cite{trianta}), with $x$ dependence for
$Q_s$ slightly different from the ansatz used here.

Embeding Eq. (\ref{sdip}) in the definition on Eq. (\ref{corel})
we obtain the following analytical expression for the color field
correlator
\begin{eqnarray}
\widetilde{C}\,(x,\rk_{\perp})= \frac{4\pi}{Q_{sA}^2(x)}\, \exp \left( -\frac{\rk_{\perp}^2}{Q_{sA}^2(x)} \right)\,,
\label{csat}
\end{eqnarray}
which also obeys the normalization  condition $\int d^2\rk\,
\widetilde{C}(x,\rk_{\perp})/(2\pi)^2=1$. In the next section, we
compute the photonuclear cross section using the correlator above
and in addition consider the complete photon flux instead of the
approximation leading to Eq. (\ref{dsdy_cgc}). The final
expression reads now  as
\begin{eqnarray}
\frac{d\sigma_{AA\rightarrow Q\overline{Q}X}}{dY}= \omega \,  \frac{dN(\omega)}{d\omega}\,\frac{\alpha_{em}e_q^2}{2\pi^2} \int\limits_{0}^{+\infty} d\rk_{\perp}^2\, \pi R_A^2\, \widetilde{C}\,(\rk_{\perp})
\left\{ 1+ \frac{4(\rk_{\perp}^2-m_Q^2)}{\rkn_{\perp}\sqrt{\rk_{\perp}^2+4m_Q^2}}\,{\rm arcth}\,\frac{\rkn_{\perp}}{\sqrt{\rk_{\perp}^2+4m_Q^2}} \right\}\!,
\label{dsdy_phen}
\end{eqnarray}
where we define the  rapidity $Y\equiv \ln(1/x)=
\ln(2\,\omega\,\gamma_L/4m_Q^2)$ and the variable transformation
between $\omega$ and $Y$ should be  carried out. The equivalent
photon flux is taken from Eq. (\ref{fluxint}).

\section{Results and discussions}
\label{discussion}

In this section we  present the numerical calculation of the
rapidity $Y$ distribution and total cross section for charm and
bottom photonuclear production. In particular, we are  focusing
mostly on LHC domain where small values of $x$ would be probed. At
RHIC, $x = (M_{Q\overline{Q}}/200 \, {\rm GeV}) e^{-y}$, which
implies  $x > 10^{-2}$ and, consequently, small deviations between
the high energy QCD approaches. In the following, one considers
the charm and bottom  masses  $m_c=1.5$ GeV and $m_b=4.5$ GeV ,
respectively. Moreover, for PbPb ($A=208$)  collisions at LHC, one
has the c.m.s. energy of the ion-ion system
$\sqrt{S_{\mathrm{NN}}}=5500$ GeV and the Lorentz factor
$\gamma_L=2930$.

\begin{figure}[t]
\begin{tabular}{cc}
\psfig{file=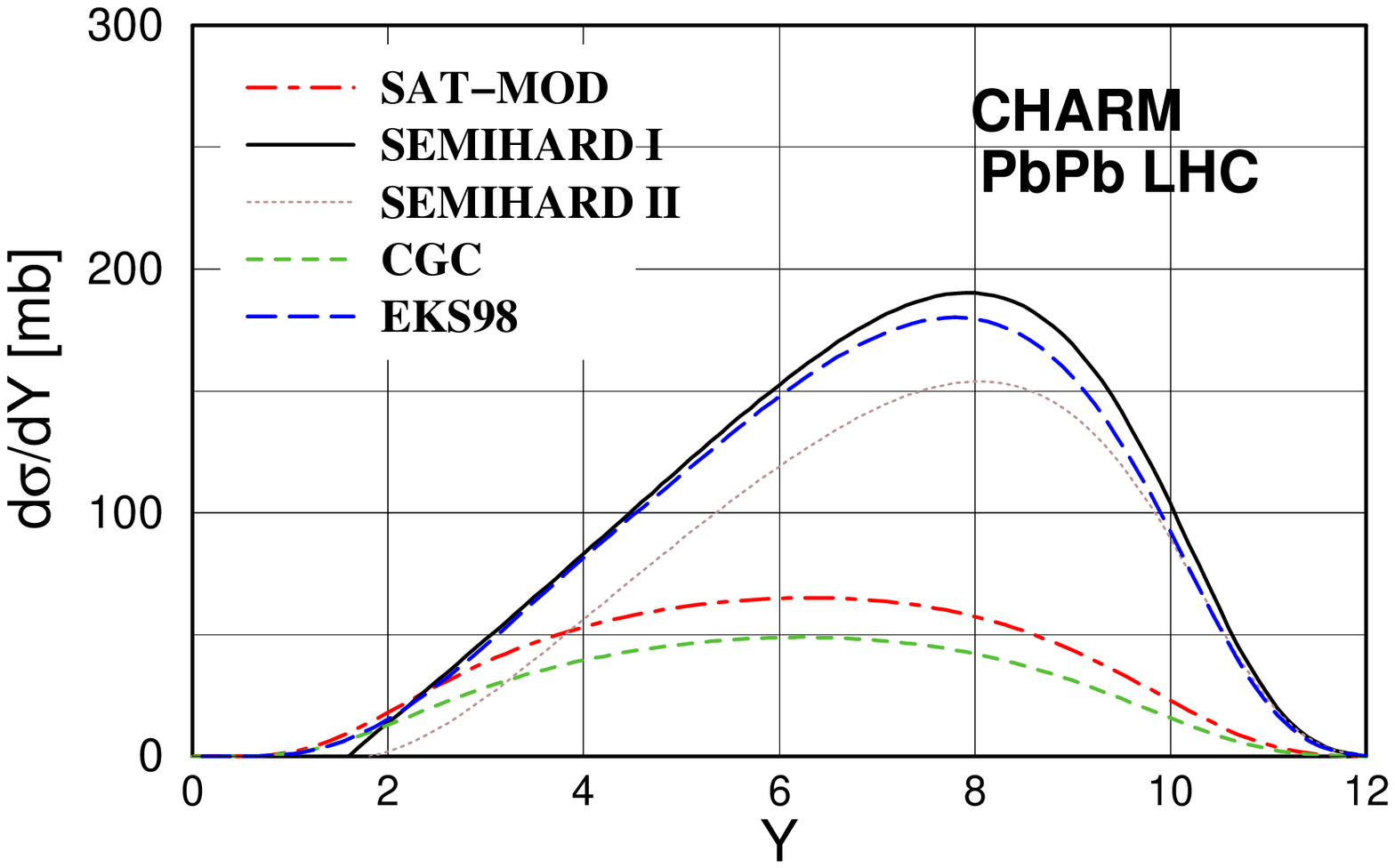,width=83mm} & \psfig{file=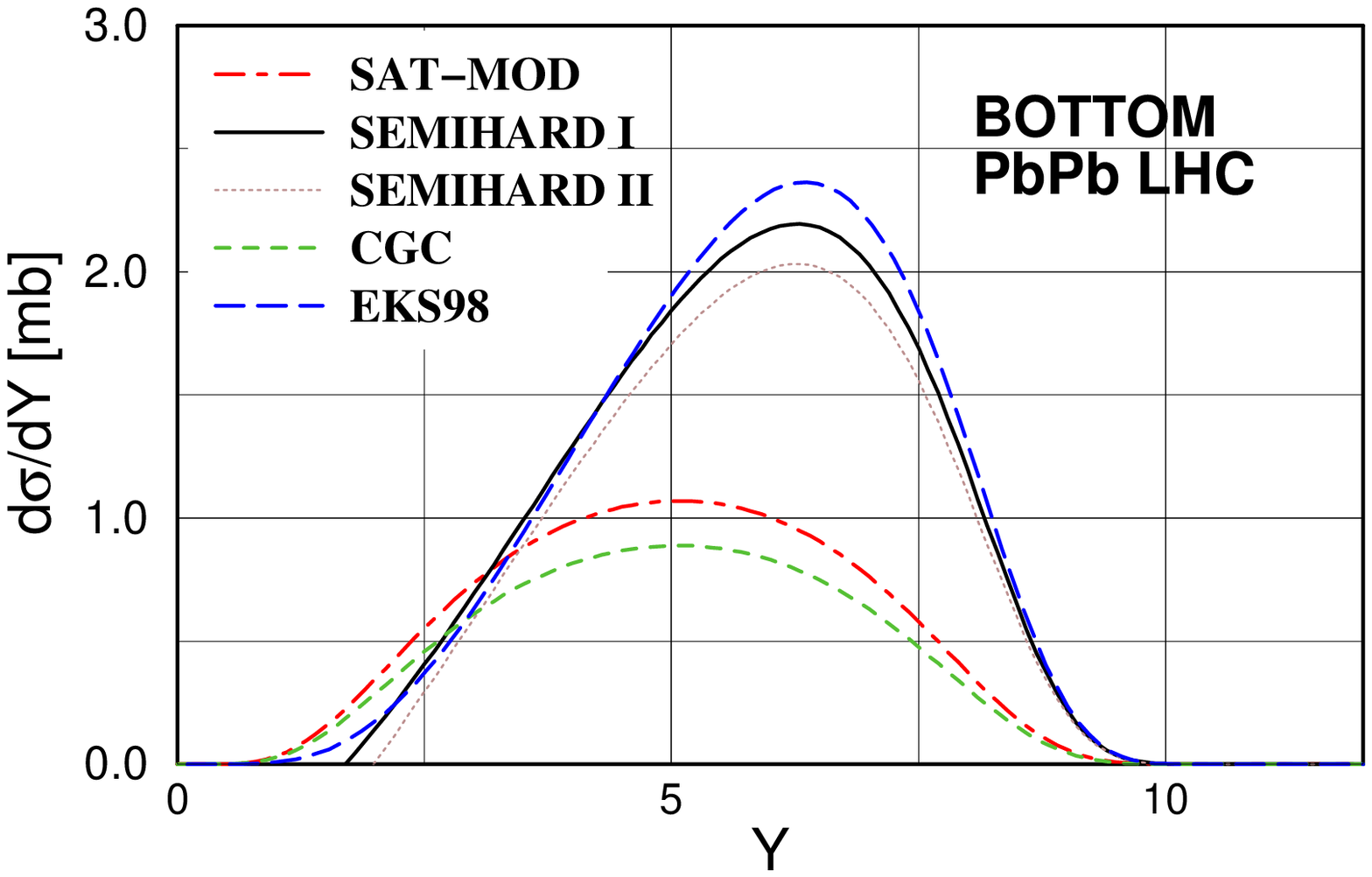,width=80mm}
\end{tabular}
\caption{\it The rapidity distribution for the distinct  high
energy approaches: collinear approach (long-dashed lines),
semihard formalism (solid and dotted lines), saturation model
(dot-dashed lines) and color glass condensate (dashed lines).  The
corresponding mass values are $m_c=1.5$ GeV and $m_b=4.5$ GeV.}
\label{fig1}
\end{figure}

In Fig. \ref{fig1} are shown the rapidity distribution,  for the
distinct high energy approaches considered before.  The collinear
result is denoted by the long-dashed curves, where has been used
Eqs. (\ref{sigAA}) and (\ref{sigpho}) employing the EKS98
parameterization for the collinear nuclear gluon function. The
solid and dotted lines label the semihard
($\rk_{\perp}$-factorization) results, where one has used Eq.
(\ref{sigmakt}) and the ansatz given by Eq. (\ref{nucugf}) for the
unintegrated gluon function. Two possibilities for the nucleon
gluon distribution are considered: (I) GRV94(LO) - solid line -
and (II) GRV98(LO) - dotted line. The saturation model results are
denoted by the dot-dashed lines, with the input given by Eq.
(\ref{sigmaphot}). The  color glass condensate prediction (dashed
lines), is given by our phenomenological ansatz using expressions
(\ref{csat}) and (\ref{dsdy_phen}). The predictions for the
collinear approach and the semihard formalism are similar for both
charm and bottom production and give somewhat larger values than
the saturation and CGC results. One possible interpretation for
the similarity between the predictions of the semihard approach
and the collinear one is  that the expected enhancement in the
$\rk_{\perp}$-factorization formalism, associated to the
resummation of the $(\alpha_s \, \ln \frac{\sqrt{s}}{m_Q})^n$ in
the coefficient function \cite{CCH}, is not sizeable for inclusive
quantities  in  the kinematic region of the future colliders. This
feature holds the trend already verified at nucleon  level as in
photon-proton and photon-nuclei collisions
\cite{Goncalves:2003zy}. Even inclusive cross sections in
hadroproduction provide similar results between collinear and
semihard (dipole) approaches \cite{Rauf}. Probably, a more
promising quantity to clarify this issue would be the transverse
momentum $\rp_{\perp}$ distribution. In this case, the semihard
approach seems to be in better agreement with experimental data in
the $pp$ collisions than the collinear approach \cite{semih}.

Our phenomenological ansatz within the CGC formalism gives similar
results as the saturation model, but should be noticed that the
physical assumptions in those models are distinct. While  Eq.
(\ref{sigmanuc}) considers multiple scattering on single nucleons,
our expression for the dipole-nucleus cross section
[Eq.(\ref{sdip})] assumes scattering on a black area filled by
partons coming from many nucleons. It is important to emphasize
that the current experimental data for the nuclear structure
function can only be described if the first  choice is implemented
\cite{armesto}. However, the correct expression for
$\sigma_{dip}^A$ in the kinematical range of the future colliders
is still an open question.

\begin{table}[t]
\begin{center}
\begin{tabular} {||c|c|c|c|c||}
\hline
\hline
$Q\overline{Q}$   & {\bf Collinear} & {\bf SAT-MOD} & {\bf SEMIHARD I (II) } &  {\bf CGC} \\
\hline
 $c\bar{c}$ & 2056 mb & 862 mb &  2079 (1679.3) mb & 633 mb \\
\hline
 $b\bar{b}$ & 20.1 mb  & 10.75 mb & 18 (15.5) mb & 8.9 mb\\
\hline
\hline
\end{tabular}
\end{center}
\caption{\it The photonuclear heavy quark  total cross sections for
ultraperipheral heavy ion collisions at LHC ($\sqrt{S_{\mathrm{NN}}}=5500$ GeV)
for PbPb.}
\label{tabhq}
\end{table}

Let us now  compute the integrated  cross section considering the
distinct models. The results are presented in Table \ref{tabhq}
for charm and bottom pair production. The collinear approach gives
a larger rate, followed by the semihard approach, a clear trend
from the distribution on rapidity. The saturation model and CGC
formalisms give similar results, including a closer ratio for
charm to  bottom production. Concerning the CGC approach, our
phenomenological educated guess  for the color field correlator
seems to produce quite reliable estimates.

Lets estimate the uncertainties present in our predictions using
the collinear  approach.  We have that for bottom production our
prediction decrease  by $\approx$ 20 $\%$ if we assume $m_b =
4.75$ GeV and by  $\approx 10\,\%$ if   we assume that
factorization scale is $\mu^2 = m_b^2$. Moreover, if the MRST
gluon density \cite{mrst} is used instead of  GRV98
parameterization, our predictions decrease by $\approx 10\,\%$.
For charm production, the differences are larger due to the small
values of $x$ probed in that process. Our results for the total
charm production cross section in the  collinear approach are
similar to those ones computed in Ref. \cite{ramonaklein}.
However,  they differ largely  for bottom production, even using
the same set of scales and parton distributions. We believe that
our  results are reliable, since our prediction for photon-nucleon
interaction is  consistent with the HERA data
\cite{Goncalves:2003zy}, as well as with the simple expectation
$\sigma_{\gamma A \rightarrow b\overline{b} X} \approx A \times
\sigma_{\gamma p \rightarrow b\overline{b} X}$.

Regarding the semihard approach we have checked  the uncertainties
coming from quark mass and different choices for the gluon
parameterization as input for the unintegrated function. In
comparison with the default value $m_c=1.5$ GeV for charm,  we
have an enhancement of $\approx 35\, \%$ using $m_c=1.2$ GeV,
whereas the result decreases the same amount for  $m_c=1.8$ GeV
(similar results stand for bottom). Considering the default value
for the quark mass, the uncertainty when using the  GRV98
parameterization is of order $\approx 20\,\%$. In Table
\ref{tabhq} we provide the cross section using (I) GRV94 and (II)
GRV98 parameterizations.  There  is an additional uncertainty
coming from the energy scale $\mu^2$ entering on the strong
coupling constant. Here we have used the optimal choice, giving
correct results at  the nucleon level and allowing simple
translation to the dipole (position space) representation, since
the energy scale does not depend on the quark transverse
momentum.

In order to check if the  differences between the saturation
approach and CGC come from the integration weights in Eqs.
(\ref{sigmakt}) and (\ref{dsdy_phen}) or from a different mass
number dependence for the saturation scales, we have used our
ansatz Eq. (\ref{sdip})  in the calculation  of the photonuclear
cross section. We have found that for lead  the CGC ansatz gives a
result $\approx 40\,\%$ lower than the saturation model, whereas
for calcium the results are almost identical. Therefore, as the
deviation between saturation model and CGC comes mostly from the
different definition for the $A$-dependence of $Q_{sA}(x)$, an
experimental analysis of this process for different nuclei can be
useful.

In conclusion, the photonuclear production of heavy quarks allow
us to constraint already in the current nuclear  accelerators the
QCD dynamics since the main features from photon-nuclei collisions
hold in the coherent ultraperipheral reactions. As a summary, we
have computed  the photonuclear production of heavy quarks in
ultraperipheral heavy ion collisions. One obtains the  integrated
cross section and the rapidity distribution through well
established QCD approaches, namely  the collinear and semihard
factorization formalisms  as well as the saturation model. For the
first time, quantitative predictions for the latter two approaches
are presented, whereas previous collinear calculations are
consistently corroborated.  Moreover,  the color glass condensate
formalism has been  considered  using a simple educated guess  for
the color field correlator in the medium, which allowed us reach
to  reliable  estimates  at LHC energies.

\section*{Acknowledgments}
M.V.T.M. thanks the support of the High Energy  Physics
Phenomenology Group at the Institute of Physics, GFPAE IF-UFRGS,
Porto Alegre. This work was partially financed by the Brazilian
funding agencies CNPq and FAPERGS.

\end{document}